\documentclass[twoside,a4paper]{article}
\usepackage{amsmath,graphicx,amssymb,fancyhdr,amsthm,,textcomp} 
\usepackage{soul} % Enables overstrike \st for removed text.
 
\usepackage[usenames]{xcolor}

\theoremstyle{definition} 
  
\theoremstyle{remark}  
  
\def\beq{\begin{eqnarray}}  
\def\eeq{\end{eqnarray}}  
\def\bsp{\begin{split}}  
\def\esp{\end{split}}

\begin{document}  
  
\title{\Large\textbf{Geometric Horizons}}  
\author{{\large\textbf{Alan A. Coley$^{1}$, David D. McNutt$^{2}$, and Andrey A. Shoom$^{3}$}}
%EndAName  
%\address{  
          \vspace{0.3cm} \\ 
                         $^{1}$ Department of Mathematics and Statistics,\\ 
                         Dalhousie University, 
                         Halifax, Nova Scotia,\\ 
                         B3H 3J5, Canada 
          \vspace{0.3cm}\\
                    $^{2}$ Faculty of Science and Technology,\\ 
                         University of Stavanger, 
                         N-4036 Stavanger, Norway         
	      \vspace{0.3cm}\\ 
	      $^{3}$ Department of Mathematics and Statistics, \\
	      Memorial University,
St. John’s, Newfoundland and Labrador,\\  A1C 5S7, Canada                         
		  \vspace{0.3cm}\\
          \texttt{$^{1}$ aac@mathstat.dal.ca, $^{2}$ david.d.mcnutt@uis.no, $^{3}$ ashoom@ualberta.ca} }

\date{\today}  
\maketitle  
\pagestyle{fancy}  
\fancyhead{} % clear all header fields  
\fancyhead[EC]{}  
\fancyhead[EL,OR]{\thepage}  
\fancyhead[OC]{}  
\fancyfoot{} % clear all footer fields  
  
\begin{abstract}   

We discuss black hole spacetimes with a geometrically defined quasi-local  horizon on which the curvature
tensor is algebraically special relative to the alignment classification.  
Based on many examples and analytical results,
we conjecture that a
spacetime horizon is always more algebraically special (in all of the
orders of specialization) than other regions of spacetime.  
Using recent results in invariant theory, such
geometric black hole horizons can be identified by the alignment
type {\bf II} or {\bf D} discriminant conditions in terms of
scalar curvature invariants, which are not dependent on spacetime
foliations.
The above conjecture is, in fact, a suite
of conjectures (isolated vs dynamical horizon; four vs higher dimensions; zeroth
order invariants vs higher order differential invariants).  
However,
we are particularly interested in applications in four dimensions and especially the
location of a black hole
in numerical computations.
  
%%%%%The question of whether these definitions and conjectures are useful, or perhaps need additional refinement in the case of dynamical horizons, will have to
%%%%%be further evaluated both analytically and numerically. 

\end{abstract} 
%%%%%%%%%%%%%%%%%%%%%%%%%%%%%%%%%%%%%%%%%%%%%

\subsection*{Introduction}

Black holes, which are exact solutions in general relativity (GR)  
(representing, for example,	 physical objects formed out of the gravitational collapse of fuel-exhausted stars), are characterised by the boundary of the region from where light can still travel to asymptotic null infinity, called the event horizon, which is usually identified as the surface of the black hole and
relates its area to the entropy.  The event horizon is essentially a
global ({\em teleological}) object, since it depends on the entire future history of the spacetime \cite{AshtekarKrishnan}.

There has been much effort to give a
general quasi-local description of a dynamical black hole
\cite{AshtekarKrishnan,Senov}.  Of
particular interest are quasi-local objects called
{\it marginally trapped tubes} (MTTs) or  {\it trapping horizons}, and the special cases of {\em dynamical horizons}  or  {\it future outer
trapping horizons} (FOTHs); in numerical work, these are also called {\em apparent horizons}.  MTTs are
hypersurfaces foliated by (closed compact space-like two-dimensional (2D) submanifolds  without boundary)
{\it marginally trapped
surfaces} (MTSs) in
which the expansion of one of the null normals vanishes and the other is
non-positive. A dynamical
horizon is a smooth 3D
submanifold of spacetime foliated
by MTSs such that the expansion of one
future-directed null normal to the foliation  vanishes,
while the expansion of the other future directed null
normal is negative 
(a FOTH has the additional condition that the directional
derivative along the second null direction is negative).

A dynamical horizon is particularly well-suited to analyze dynamical processes involving black holes,
such as black hole growth and coalescence.  
The area of a dynamical horizon necessarily increases with time \cite{AshtekarKrishnan}.  
An explicit example of a dynamical horizon is given by the Vaidya spacetime which admits
spherically symmetric MTSs \cite{Vaidya, BonnorVaidya}. For a
given mass function, the Vaidya spacetime also provides explicit examples of the
transition from the dynamical to isolated horizons.  
If a hypersurface admits a dynamical horizon structure, it is unique.
However, because a spacetime may have several distinct black holes, it may admit several
distinct dynamical horizons.  For dynamical horizons which are also FOTHs, two
non-intersecting horizons generally either coincide or one is contained in the other \cite{AshtekarKrishnan}.

It is believed that closed
MTSs constitute an important
ingredient in the formation of black holes, which
motivates the idea of using MTTs 
as viable replacements 
for the event horizon
of black holes \cite{Senov}.
Unfortunately, since the
2D apparent horizons depend on the choice
of a reference foliation of spacelike hypersurfaces, MTTs and consequently  
trapping horizons and dynamical horizons are highly
non-unique \cite{AG2005}.  
There have been some attempts to provide a physically sound criterion for selecting a preferred MTT such as, for example,
in which the shear scalars along ingoing/outgoing null directions foliated by 2D spacelike surfaces vanish \cite{Umbilical}.

Realistic black holes interact with their environment and are consequently
dynamical.  The gravitational collapse leading to black hole formation is
also a highly dynamical process.  
It is crucial to locate a black hole locally, which may not rely on the existence of
an event horizon alone.
A significant fraction of
research in numerical relativity aims at predicting with high precision
the waveforms of gravitational waves generated in the merger of
compact-object binary systems or in stellar collapse to form black holes.
Comparison with templates played a crucial role in the recent
observations of gravitational waves from black hole mergers  by the LIGO collaboration \cite{LIGO}.

In numerical studies of time-dependent collapse, it is often more practical to track apparent
horizons or trapping
horizons \cite{Booth2005}.
In contrast with the event horizon, which is a
global concept defined using the global structure of spacetime, the
apparent horizon is a quasi-local concept and is intrinsically
foliation-dependent.
In this paper we propose a foliation invariant and more geometrical approach, which is possible due 
to recent results in invariant theory.

\newpage

\subsection*{Scalar polynomial curvature invariants}

The algebraic classification of
the Weyl tensor and the Ricci tensor in arbitrary dimensions using the boost weight decomposition \cite{class} can 
be refined utilizing the restricted eigenvector
and eigenvalue structure of their associated curvature operators \cite{BIVECTOR}, allowing for necessary conditions to be defined for a particular algebraic type in terms of a set of
discriminants.
A {\em scalar polynomial curvature invariant of order $k$} 
(or, in short, a scalar polynomial invariant or $SPI$) is a scalar obtained by
contraction from a polynomial in the Riemann tensor and its
covariant derivatives up to the order $k$. Black hole spacetimes are completely characterized 
by their $SPI$s \cite{inv}.
In particular, we can use discriminants to study the necessary conditions in arbitrary dimensions, in terms of simple
$SPI$s,  allowing for the algebraic classification of the higher dimensional Weyl and Ricci tensor when treated as curvature operators, for the spacetime  to be of algebraic type {\bf II} or {\bf D} \cite{CH}.

For example,  in 5D
the necessary condition 
for the trace-free Ricci tensor, $S_{ab} = R_{ab} - \frac{1}{5} R g_{ab}$, to be of algebraic type {\bf II}/{\bf D} 
is that the discriminant ($SPI$) ${}_S^5D_5$ is zero, 
and the necessary conditions 
for the Weyl tensor to be of type {\bf II}/{\bf D} 
is that the  $SPI$s 
${}_W^{10}D_{i}$ ($i=8,9,10$) vanish \cite{CH}.
As an illustration, the 5D rotating black ring \cite{RBR} is generally of type ${\bf I_i}$,
but on the horizon the discriminant ${}_T^5D_5$ 
of the trace-free part of the operator 
$ T^a_{~b} = C^{acde}C_{bcde}$ vanishes (and  ${}_T^5D_4>0$), 
which signals that the spacetime is of Weyl type {\bf II} 
on the horizon.

We are primarily interested in the 4D case here.
The necessary type {\bf  II}/{\bf D} discriminant condition ${^4}D_4=0$ for 
the trace-free 
(${\it s_1}=0$) symmetric Ricci tensor $S$ in 4D is \cite{CH}:
\begin{equation}
%\label{mathcalD}
\mathcal{D} \equiv {^4}D_4 = -s_3^2(4 s_2^3 - 144 s_2 s_4 + 27s_3^2) + s_4(16 s_2^4 - 128 s_4 s_2^2  + 256 s_4^2) = 0,
\label{rictypeii2}
\end{equation}
where
\beq s_{2} \equiv -\frac12 S^a_{~b} S^b_{~a},~~ s_{3} \equiv -\frac13 S^a_{~b} S^b_{~c} S^c_{~a}, ~~
s_{4} \equiv \frac14 (2s_{2}^2 - S^a_{~b} S^b_{~c} S^c_{~d} S^d_{~a}). \label{riccidef}  \eeq
\noindent Similar conditions hold for any trace-free symmetric tensor
$T_{ab}$.

The necessary real conditions for the Weyl tensor to be of type {\bf  II}/{\bf D} are \cite{CH}:
\begin{equation}
\mathcal{W}_1 \equiv -11 W_{2}^3 + 33 W_2 W_4 - 18 W_6 = 0, \label{weyl1}
\end{equation}
\begin{equation}
\mathcal{W}_2 \equiv (W_{2}^2 - 2 W_4)(W_{2}^2 + W_4)^2 + 18 W_3^2(6 W_6 - 2 W_{3}^2 -
9 W_{2} W_4 + 3 W_{2}^3) = 0, \label{weyl2}
\end{equation}
where
\begin{equation}
W_2 = \frac{1}{8}C_{abcd}C^{abcd},
~W_3  = \frac{1}{16}C_{abcd}C^{cd}_{~~ pq}C^{pqab},
~W_4 = \frac{1}{32}C_{abcd}C^{cd}_{~~pq}C^{pq}_{~~r s}C^{rsab}, \nonumber
\end{equation}
\begin{equation}
W_6 = \frac{1}{128}C_{abcd}C^{cd}_{~~pq}C^{pq}_{~~r s}C^{rs}_{~~tu}C^{tu}_{~~vw}C^{vwab}.
\label{weyldef}
\end{equation}
These 2 real conditions are equivalent to the real and imaginary parts of the complex syzygy $I^3-27J^2=0$ in terms of
the complex Weyl tensor in the Newman-Penrose (NP) formalism \cite{kramer}.

Alternatively we can use the discriminant analysis to provide the type {\bf II/D} syzgies expressed in terms of the
$SPI$s  $W_i$ ( $i \in [1,6]$, defined above) by treating the Weyl tensor as a trace-free operator acting on the 6-dimensional vector space of
bivectors \cite{CH} 
(however, these conditions are very large). 
More practical {\it necessary} conditions can be obtained by
considering the trace-free symmetric operator
$C_{abcd}C^{ebcd} - 2 W_2 \delta_a^{~e}$;
applying the discriminant analysis we find the coefficients of the characteristic equation are 
$w_{2} = 8(W_2^{~2} - 2 W_4)$
(and similarly for $w_{3}, w_{4}$) 
and so the necessary condition for this operator to be type {\bf II/D} is given by equation \eqref{rictypeii2}
(with the $s_i$ replaced by $w_i$).

The alignment classification can be applied to any rank tensor. To consider whether the
covariant derivatives of the Ricci tensor, $R_{ab;cd...}$,  are
of type {\bf II} or  {\bf D}, we can use the eigenvalue structure of the operators associated with the derivatives of the Ricci curvature and impose the type {\bf II}/{\bf D} necessary conditions. 
This can be repeated for the Weyl tensor and in arbitrary dimensions \cite{invhigher}.
For example, for the covariant derivative of the Weyl tensor,
$C_{abcd;e}$, in 4D we
can consider the second order symmetric and trace-free
operator ${^1}T^a_{~b}$ defined by:

\begin{equation}
{^1}T^{a}_{~b} \equiv C^{cdef;a}C_{cdef;b} - \frac{1}{4} \delta^{a}_{~b} {^1}I_2 
\end{equation}
where ${^1}I_2\equiv
C^{abcd;e}C_{abcd;e}$,
and we have the corresponding 4th, 6th, 8th order invariants
${^1}I_4,
{^1}I_6,
{^1}I_8$.
Computing the coefficients of the characteristic equation we obtain 
${^1}s_{2} = -\frac12 {^1}I_4 + \frac18 {^1}I_2^2$ (and similarly for
${^1}s_{3}, {^1}s_{4}$).
The necessary condition for this operator to be of type {\bf II/D} (of the form $\mathcal{D} \equiv {^4}D_4=0$)
is equivalent in form to the condition given 
in equation \eqref{rictypeii2} with $s_{i}$ replaced by   ${^1}s_{i}$, and can be expanded out explicitly.
For example, for the operator
${^1}T^{a}_{~b}$ defined above for the type {\bf D} Kerr metric,
the vanishing of
${}_T^4D_4$ implies that $C_{abcd;e}$ is of type {\bf D}/{\bf II} 
on the horizon.

It is known that differential invariants, constructed  from $SPI$s
in terms of the Weyl tensor and its covariant derivatives,
detect horizons for several type
{\bf D} stationary solutions \cite{AbdelqaderLake2015}.
The construction of the class of Page-Shoom $SPI$s that detect the horizons of stationary black holes
exploits the fact that on the horizon the timelike Killing vector becomes null and is, in fact, a generator of the
horizon \cite{PageShoom2015}.  Noting that stationary horizons are a special case of weakly isolated horizons, 
the type {\bf II/D}  $SPI$s arising from the
discriminant analysis of $C_{abcd;e}$ vanish on the horizon, and it can be explicitly shown that the type {\bf II/D} discriminants
share common zeros with the Page-Shoom 
invariant ${\bf W}$ for the Kerr spacetime \cite{ADA}. Similar results using Cartan invariants are possible \cite{GANG}.

\subsection*{Examples and motivation}

There are many examples 
(some briefly discussed in this paper, but see also \cite{GANG}) that 
support the geometric conjectures to follow. 
Indeed, all of the known exact 
black hole solutions are algebraically special of Weyl (curvature) type {\bf II/D}
on the horizon \cite{Pelavas}. This 
led to an earlier conjecture that asserted that
stationary black holes in higher dimensions, possibly with the additional conditions of vacuum or asymptotic flatness, must be of Weyl type {\bf D}  \cite{Pelavas}.

There is also motivation for the conjectures
from analytical results.
Quasi-local isolated horizons, which account for equilibrium states of
black holes and cover all essential local features of event horizons
\cite{AshtekarKrishnan}, are essentially defined as a 3D 
null surface (submanifold) with topology $S_2 \times R$ with an
outgoing expansion rate which vanishes on the horizon.
It follows that the null normal vector is a local time-translational Killing vector field on the horizon,
and requires
neither asymptotic structures nor foliations of spacetime.
Every such Killing horizon  is an isolated horizon \cite{AshtekarKrishnan}.
In particular, this implies that the event horizon of Kerr geometry is
an isolated horizon. However, in general,
spacetimes with isolated horizons need not admit any Killing vector fields even
in a neighborhood.
In \cite{Isenberg} it was proven that if a stationary, real analytic, asymptotically flat
vacuum black hole spacetime of dimension $D \geq 4$ contains a
non-degenerate horizon with compact cross sections that are transverse
to the stationarity generating Killing vector field then, for each
connected component of the black hole's horizon, there is a Killing
field which is tangent to the generators of the horizon.

In the 4D case, and assuming the ``mild energy condition'' implied by the dominant energy condition, the existence of an induced degenerate metric tensor which locally acts as a metric tensor on the 2D tangent space, and the induced covariant derivative, 
which constitute the
geometry of a nonexpanding null surface, was demonstrated;
this then leads to the conditions that on the non-expanding weakly isolated  horizon
the Ricci and Weyl tensors are of type {\bf II/D} \cite{Ashtekar}.
This local result has been generalized to non-expanding null surfaces in arbitrary dimensions (and the result is applicable to surfaces of any topology); 
indeed, it was shown that if the expansion of a null surface vanishes, then the shear must also vanish and a covariant derivative can be induced on each non-expanding null surface \cite{Lewandowski}.  
It can also be shown \cite{AD}  that
for any weakly-isolated horizon the Riemann tensor and the covariant derivatives of the Riemann tensor 
are of type {\bf II} on the horizon.

We note that when a star collapses to
form a black hole, the exterior of the black hole eventually settles down to a
stationary state, most likely described by the Kerr metric. Despite what the interior of the
black hole settles down to, this leads by continuity to the
expectation that there will be a region of the interior near the horizon that
should be close to the interior Kerr metric.  Inside the black hole event
horizon the Kerr metric has an inner horizon which is also a null surface.
However, the inner horizon is unstable, so for a spacetime that begins close
to the Kerr metric, the inner horizon should be replaced by 
something else, perhaps even a singularity \cite{MassInflation, PoissonIsrael, Ori, MarolfOri}.
There are a variety of analytic arguments, mathematical results, and numerical
simulations that indicate that this singularity maintains the inner horizon's
character as a null surface \cite{AshtekarKrishnan,Ashtekar}.

This supports the notion that the horizon is smooth and unique at later times
and, in principle, can be identified by algebraic/geometrical conditions. It is possible that as
we follow this unique, smooth surface back in time (during the physics of collapse or merger),
this surface suffers a bifurcation and this surface is no longer unique or
smooth (or even differentiable). But it is plausible that there exists a unique, smooth
geometric horizon that shields all other horizons (or at least identifies the region of
interest).

\subsection*{The Geometric Horizon  Conjecture}

This consequently motivates us to 
conjecture that there is a geometrically defined                                                                                                         
unique, locally determinable, smooth (dynamical) horizon
on which the curvature tensor is algebraically special.
In particular, this implies that a spacetime horizon is always more algebraically special
(in all of the orders of specialization) than (all) other regions of spacetime. 
Such geometric black hole horizons can be identified and located by the type {\bf II/D} conditions
in terms of  $SPI$s, 
which are 
not dependent on spacetime foliations. To state the conjectures, we will say a tensor ${\bf T}$ is {\em $n^{th}$-order algebraically special} if ${\bf T}$ and all covariant derivatives of ${\bf T}$ up to order $n$ are of algebraic type {\bf II} or more special.

\paragraph{Conjecture Part I:}
{\em{If a black hole spacetime is zeroth-order algebraically general, then on the geometric horizon the spacetime
is algebraically special. We can identify this geometric horizon using scalar curvature invariants.}}

This is the more practical part of the Conjecture and will hopefully be
useful to numerical relativists who study the collapse or merger of real black holes, which are
generically of general algebraic type away from the horizon.
The conjecture might be qualitatively different for isolated and dynamical horizons.   
The issue then becomes one of finding effective ways to do computations.

\paragraph{Conjecture Part II:}
{\em{If a black hole spacetime is zeroth-order algebraically special (and on the horizon the spacetime
is thus also
algebraically special), then if the black hole spacetime is first-order algebraically general, 
then on the horizon the spacetime is first-order
algebraically special.
We can identify this  geometric horizon using differential scalar curvature invariants.}}

If necessary, this can be repeated for higher order covariant derivatives. 
This is the more theoretical and analytical part of the Conjecture,
and can be applied to exact solutions. 
Note that in general we may not wish that the covariant derivatives be algebraically special 
(i.e., of type  {\bf II}/{\bf D}) to each order (i.e., of type $D^k$) on the black hole horizon, 
as this might be too restrictive. 

\paragraph{Comments:}      
The algebraic conditions expressed in terms of $SPI$s essentially define a {\em{geometric horizon}}.
In order to make the definition more precise, we need to focus on physical black hole solutions  
(both exact black hole solutions
and generic physical collapse and black hole coalescences)
and in order to prove
definitive results we need to append some 
physical conditions to the definition such as, for example,  energy conditions, a particular theory of gravity
(e.g., we assume GR; in principle some conditions may be different for different theories), 
and perhaps other asymptotic conditions.
We also note that $SPI$s may not specify the geometric horizon completely in the sense that they may also vanish at fixed points
of any isometries and along any axes of symmetry. However, we expect that the identification of a smooth surface for
physical situtations is always possible. Unlike 
apparent horizons, a geometric horizon does not depend on a chosen foliation in the
spacetime.

Although this conjecture is also intended to apply in higher dimensions \cite{GANG},
we are primarily interested in applications in 4D, and particularly
in numerical computations.
Indeed, the above conjectures are, in fact, a suite of conjectures (isolated vs dynamical horizon; 4D vs higher dimensions;
zeroth order invariants vs higher order differential invariants). In physical problems with dynamical evolution the 
horizon might not be unique, or may not exist at all, and amendments to the conjecture may be necessary
(e.g., it may be appropriate to replace the vanishing of invariants in the definition of a 
geometric horizon as an algebraically special hypersurface, with the conditions that the                            
magnitudes of certain $SPI$s take their smallest values).

In a sense the conjecture refers to ``peeling properties'' (of the geometrical curvature)
close to the horizon. The 
curvature is of algebraically special type {\bf II} close to the horizon but it is plausible that, as gravitational 
wave modes (of algebraic types {\bf III} and {\bf  N}) fall off
more quickly to infinity, the {horizon
eventually settles down to be type {\bf D}} under some reasonable asymptotic conditions.

\subsection*{Discussion}

The question of whether these definitions and conjectures are useful 
will have to be further evaluated. Although we have attempted to
support the conjectures with  analytical results and practical examples (also see \cite{AD}), 
further work is required and perhaps 
additional refinement of the conjectures will be necessary.
In particular, it is of primary importance to study geometric horizons  numerically
in 4D in physically relevant asymmetric collapse and black hole coalescences.

\paragraph{Dynamical horizons:}
The conjecture is intended to apply to dynamical horizons.     
It is much more difficult to study dynamical horizons, but let us discuss some preliminary encouraging
results \cite{AD}.

Let us first consider the imploding spherically symmetric metric in advanced coordinates \cite{kramer}:

\beq ds^2 = - e^{2\beta(v,r)} \left(1 - \frac{2m(v,r)}{r}\right) dv^2 + 2 e^{\beta(v,r)}dv dr + r^2 d \Omega^2 \nonumber \eeq

\noindent where $m(v,r)$ is the mass function and $\beta(v,r)$ is an arbitrary function. 
It is known that the unique spherically symmetric FOTH is given by the surface $r-2m(v,r) = 0$, which is
expansion-free. 

Using the NP formalism, the Riemann tensor for any spherically 
symmetric black hole
solution is found to be of type {\bf II/D} on the horizon; i.e., at the algebraic level the
Ricci tensor always detects the horizon \cite{AD}.  However, the covariant derivatives of Ricci and Weyl will 
generally be of type {\bf I}.  Note that for this class of metrics the Ricci tensor cannot be globally type {\bf D} unless
$m_{,v}=0$, in which case the horizon is isolated.

Let us assume that $m_{,v} \neq 0$ and consider dynamical horizons.
Since no field equations have been imposed, we need additional conditions.    
For the class of spherically symmetric metrics admitting a geodesic-lined horizon which is a shear-free MTT \cite{Senov}, with the additional condition that the
frame vectors normal to the surface are geodesics, requires that the metric function $\beta$ satisfies 
$\beta_{,v} = 0$ \cite{AD}. It can then be shown that
the Riemann tensor and its covariant derivatives for any spherically symmetric 
black hole solution with $\beta_{,v} = 0$ are of type {\bf II/D} on the horizon \cite{AD}.
Hence the imploding exact Vaidya solution and the class of exact Tolman-Bondi solutions admitting MTTs \cite{Booth2005}  
satisfy the type {\bf II/D} conditions on the horizon.

We next consider the Kastor-Traschen dynamical two-black-hole solution \cite{KT}, 
which describes two
charge-equal-to-mass black holes in a spacetime with a positive
cosmological constant.  The spatial part of the metric is
written in coordinates centered at each of the black hole positions ($r_i=0$, $i=1,2$) and represents
a 3D infinite cylinder with 2D cross-sectional area of $4\pi m_i^2$, where
$m_1$ and $m_2$ are the black hole masses.
When the sum of the black hole masses does not exceed a critical mass, the
black holes coalesce and form a larger black hole.

At earliest times  ${\cal W}_1 \to  0$ as 
$\tau\to-\infty$ (in these spacetimes $\tau\in(-\infty,0)$), and
there are two 3D geometric horizons enclosing the 2 black holes.
It is found \cite{ADA} that ${\cal W}_2$  is identically zero, and  
the type  {\bf II}/{\bf D} discriminant  ${\cal W}_1$ vanishes
on segments of the symmetry-axis, at  the
black hole coordinate locations $r_i=0$ 
and on an additional ``dynamical" 2D (cylindrical) surface  around the
symmetry-axis (this 2D surface appears concurrently with the two 3D black hole horizons in the center of
mass plane).
At the earlier stages of the coalescence the 2D  surface has
a finite cross-sectional
radius (from the symmetry-axis), but at later stages this surface 
expands as the two black holes move toward one another.
There is a measure of the black hole separation that can be introduced
such that as $\tau \to 0^- $ this measure
approaches zero as the two black holes merge \cite{ADA} and the 2D surface forms around the two black
holes,  so that it is possible to identify the
location of a geometric horizon in the dynamical regime.
As $\tau\to0^-$,  ${\cal W}_1 \to  0$
and in
the quasi-stationary regime there will be a single 3D horizon;
after merger the spacetime eventually
settles down to a type {\bf{D}} Reissner-Nordstrom-de Sitter black hole 
of mass $m_1+m_2$ (which is known to have a 3D geometric horizon \cite{GANG}).

In addition, it is found that there are 3D surfaces located at a finite distance from the
axis of symmetry
for which the type {\bf{ II/D}} discriminant for the trace-less Ricci tensor vanishes, and
where the Ricci tensor is consequently of type {\bf{ II/D}}.
There is also numerical
evidence for a minimal 3D geometric surface evolving in time where the  invariant
${\cal W}_1$ locally takes on a constant non-zero minimum value.
These results are suggestive and lend support for the existence of a geometric horizon in
the dynamical regime in these exact spacetimes, but further analysis is necessary which will be presented 
in the future.

%%%%%%%%%%%%%%%%%%%%%%%%%%%%%%%%%%%%%%
%%%%[[The proper distance between the black holes diverges logarithmically during all stages of the evolution
%%%%$\tau\in(-\infty,0)$.  In addition, the proper distance between any spacetime event separated by a space-like
%%%%interval from either of the black holes diverges logarithmically as well]].  [[At the earlier stages of the
%%%%coalescence the surface has finite coordinate and proper distances ($\tau=const$), i.e., its cross-sectional radius,
%%%%from the symmetry $z$-axis.  At the later stages the surface expands (and moves away form the BHs?)and in the limit
%%%%$\tau\to0^-$ the proper distances which measure the radius diverge.  ]] [[ANDREY says NO:  The fact that the 2D
%%%%surface expands to asymptotic infinity while an apparent horizon of the Kastor-Traschen two-black-hole solution is
%%%%located at a finite coordinate distance from the axis of symmetry (suggests that the geometric horizon generally does
%%%%not detect the location of the apparent horizon.)]]

\paragraph{Computability:}

The calculation of gravitational wave signals in the theoretical modelling of 4D
sources in the framework of GR is well understood
\cite{T}. In higher dimensions, numerical simulations of rapidly spinning objects have been studied \cite{cook}.
Of course, if our ultimate aim is to provide potentially useful results for                                           
numerical relativists, computability is an important issue. In                                          
this regard using  Cartan invariants in the NP  approach is certainly an advantage.                                                      
Indeed, we have already used the NP approach to address certain problems, proving its utility, 
and we aim to develop this further in future research \cite{AD}.
As an illustration, in the 4D Kerr-NUT-AdS metric \cite{kramer}
the cohomogeneity is 2D, 
and the Page-Shoom invariant 
${\bf W}$ produces a degree eight, 
first order $SPI$ that detects the horizon \cite{PageShoom2015}.  
On the other hand, the two NP spin coefficients (Cartan scalars)
$\rho$ and $\mu$, vanish on the event horizon,  which also implies that the first covariant derivative of the Weyl spinor is of type {\bf D} on the event horizon. These Cartan invariants are easier to compute than the related SPIs \cite{GANG}.

\newpage

\section*{Acknowledgements}  
 
This work was supported through NSERC (A. A. C.), NSERC Discovery
Grant 261429-2013 (A. A. S.), the AARMS collaborative research group (A. A. S.), and the Research Council of Norway, Toppforsk grant no. 250367: Pseudo-Riemannian Geometry and Polynomial Curvature Invariants: Classification, Characterisation and Applications (D. D. M.). GRTensorII was used in some of our calculations. We also would like to thank Ivan Booth, Kayll Lake and Malcolm MacCallum for discussions at the early stages of this work.

\end{document}